\documentclass[11pt,a4paper]{article}
\pdfoutput=1
\usepackage{jheppub}

\usepackage{color}
\usepackage{amsmath}
\usepackage{pifont}   
\usepackage{verbatim}       
\usepackage{acronym} 

\usepackage{amsfonts}
\usepackage{amssymb}
\usepackage{mathrsfs} 
\usepackage{graphicx}
\usepackage{multirow} 
\usepackage{slashed} 
\usepackage{array}

\usepackage{graphicx}
\usepackage{epsfig}
\usepackage{lscape}
\usepackage{rotating}
\usepackage{epsf}
\usepackage{bm}
\usepackage{booktabs}
\usepackage{array}
\usepackage{caption}
\usepackage{subcaption}
\usepackage[utf8]{inputenc}
\usepackage{float}

\usepackage{multirow}
\usepackage{array,booktabs,colortbl,multirow}
\usepackage{colortbl,xcolor,colordvi,color}
\usepackage{rotating}
\allowdisplaybreaks

\newcommand{\cmrule}{\midrule[0.25mm]}

\title{Event-level variables for semivisible jets using anomalous jet tagging}

\author[a, b]{Hugues Beauchesne}
\author[c]{and Giovanni Grilli di Cortona}

\affiliation[a]{Department of Physics, Ben-Gurion University, \\Beer-Sheva 8410501, Israel}
\affiliation[b]{Physics Division, National Center for Theoretical Sciences,\\ Taipei 10617, Taiwan}
\affiliation[c]{Istituto Nazionale di Fisica Nucleare, Laboratori Nazionali di Frascati, \\C.P. 13, 00044 Frascati, Italy}

\emailAdd{beauchesneh@phys.ncts.ntu.edu.tw}
\emailAdd{grillidc@lnf.infn.it}

\abstract{Semivisible jets are a characteristic signature of many confining dark sectors and consist of jets of visible hadrons intermixed with invisible stable particles. Since their initial proposal, considerable progress has been made in developing techniques for tagging anomalous jets. In this paper, we show that the ability to tag semivisible jets can be used to define new event-level variables which use generally ignored kinematic information and can considerably increase our ability to discover dark confining sectors. In practice, our best results are obtained by using the coefficients of the decomposition of the missing transverse momentum in terms of the transverse momenta of the jets tagged as anomalous. A benchmark scenario is introduced and the increase in significance due to these new variables is studied over some of its parameter space.}

\begin{document}

\maketitle

\section{Introduction}\label{Sec:Introduction}
Many extensions of the Standard Model (SM) contain a dark sector that confines at energies not so far removed from the QCD scale. These confining sectors are common in various models of dark matter such as Asymmetric Dark Matter \cite{Petraki:2013wwa, Kaplan:2009ag, Zurek:2013wia} and dark pions \cite{Buckley:2012ky, Hochberg:2014dra, Heikinheimo:2014xza, Freytsis:2016dgf, Okawa:2016wrr, Beauchesne:2018myj,Bernreuther:2019pfb, Beauchesne:2019ato}. They are also ubiquitous amongst solutions to the hierarchy problem such as the Twin Higgs \cite{Chacko:2005pe, Barbieri:2005ri}, Folded SUSY \cite{Burdman:2006tz}, the Hyperbolic Higgs \cite{Cohen:2018mgv} and certain variations of the Relaxion \cite{Graham:2015cka}. Other notable examples include Refs.~\cite{Strassler:2006qa, Strassler:2006im}.

One signature characteristic of many confining dark sectors is so-called semivisible jets \cite{Cohen:2015toa}. Assume a particle charged under a new confining group and that is not charged under any gauge group of the Standard Model. We will refer to such a particle as a dark quark. Given a sufficiently strong portal to the Standard Model, it could be possible to produce dark quarks at colliders. These would then go through a process of dark hadronization, creating a spray of dark hadrons. Assume some of these dark hadrons are stable and some decay to hadrons. The unstable ones would form a jet, while the stable ones would escape the detector and contribute to the missing transverse momentum $\mathbf{p}^T_{\text{miss}}$. The resulting object would be a jet with a parallel contribution to $\mathbf{p}^T_{\text{miss}}$ and is known as a semivisible jet. Previous studies of semivisible jets include the theoretical works of Refs.~\cite{Cohen:2015toa, Cohen:2017pzm, Beauchesne:2017yhh, Kar:2020bws} and the experimental search of Ref.~\cite{CMS-PAS-EXO-19-020}.

The special nature of semivisible jets leads to events that display very peculiar relations between the direction of the jets and the azimuthal direction of $\mathbf{p}^T_{\text{miss}}$. This can in principle be used to define new event-level variables to suppress the background, which has previously been done in Refs.~\cite{Cohen:2015toa, Cohen:2017pzm, Beauchesne:2017yhh}. These works however all have in common that they did not consider the possibility of tagging semivisible jets. Over the past few years, there has been considerable progress in developing tagging techniques for jets originating from new confining sectors \cite{Aguilar-Saavedra:2017rzt, Park:2017rfb, Farina:2018fyg, Heimel:2018mkt, Cohen:2020afv, Bernreuther:2020vhm, Kar:2020bws}. This therefore opens the possibility of tagging semivisible jets and using the resulting information to define new event-level variables, which could potentially increase our ability to discover dark confining sectors.

With this context in mind, the goal of this paper is to design event-level variables to discover new confining dark sectors using the ability to tag semivisible jets. Designing a full search for semivisible jets is beyond the goal of this paper, but we aim to show that such variables could considerably increase the reach of future searches.

In the end, our best results are obtained by using the coefficients of the decomposition of $\mathbf{p}^T_{\text{miss}}$ in terms of the transverse momenta of the jets tagged as semivisible jets. These two variables display sharp peaks at known values and are mostly uncorrelated. Cutting on these variables, the significance of an excess can sometimes be increased by almost an order of magnitude.

The paper is organized as follows. A benchmark scenario is first defined and the different backgrounds and simulation details are presented. Multiple potential event-level variables are then introduced. Afterward, the candidate variables are compared and a scan of the increased significance is performed over a sizable region of the parameter space of the benchmark. The possibility of imperfect tagging is taken into account. Finally, some concluding remarks are included on the applicability of these variables to other scenarios and potential issues.

\section{Benchmark, backgrounds and simulation details}\label{Sec:BBSD}
We begin this paper by introducing the benchmark scenario and discussing the relevant backgrounds and simulation details.

\begin{figure}[t!]
\centering
\includegraphics[width=8cm]{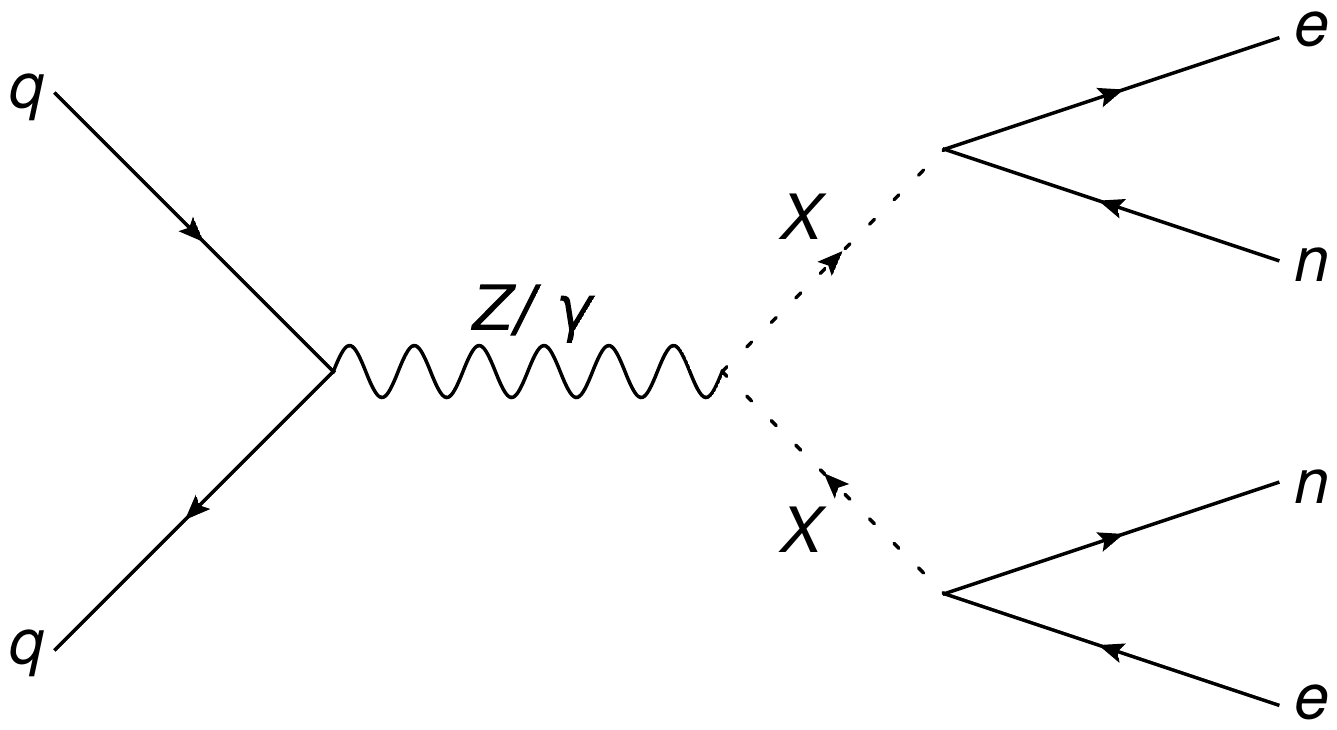}
\caption{Leading order mediator production mechanism and subsequent decay.}\label{fig:Production}
\end{figure}

Our benchmark is taken from Ref.~\cite{Beauchesne:2017yhh}, albeit the notation can be simplified for our current purposes. Assume a new confining group $\mathcal{G}$. Introduce a Dirac fermion $n$ that is neutral under the SM gauge groups and a fundamental of $\mathcal{G}$. It is a dark quark. Introduce a complex scalar $X$ that is an antifundamental of $\mathcal{G}$ and has the same SM gauge numbers as a right-handed lepton. This scalar acts as a mediator between the SM and dark sector. As long as it is not too heavy, the mediator $X$ can be pair-produced at hadron colliders via electroweak processes. In addition, one can write the Lagrangian
\begin{equation}\label{eq:LagrangianDecay}
  \mathcal{L} = \lambda_i X^\dagger \bar{n} P_R E_i + \text{h.c.},
\end{equation}
where $E_i$ are the SM leptons. This term allows the mediator $X$ to decay to a lepton and a dark quark, which itself leads to a semivisible jet.\footnote{If communications between the SM and dark sector are only via the mediator $X$, the unstable dark hadrons might not necessarily decay hadronically. It is however easy to imagine that another mediator could cause them to decay mostly hadronically.} We will assume throughout the paper that the only non-negligible $\lambda_i$ is the one corresponding to the electron. This minimizes potential flavour issues and insures that $X$ decays to $n$ and an electron. The full process at leading order is shown in Fig.~\ref{fig:Production}. The resulting signature is then two electrons and two semivisible jets. This benchmark is chosen for its simple and clean nature, but the discussion of this paper will also apply to more general signals.

The signal of the benchmark shares many similarities with leptoquark pair-production with the leptoquarks decaying to first generation leptons, albeit with a much smaller cross section. Searches for such leptoquarks typically have $t\bar{t}$ and $Z/W$ + jets as their dominant backgrounds (see for example Refs.~\cite{Sirunyan:2018ryt, Aad:2020iuy}). Which of these two backgrounds dominates depends on the details of the analysis, especially if $b$-tagged jets are vetoed or required. For dark jets that are expected to differ sufficiently from regular jets, a realistic search would require the presence of jets tagged as anomalous. Whether the dominant background is $t\bar{t}$ or $Z/W$ + jets would then depend on whether $b$-jets or QCD jets are more likely to be tagged as anomalous jets. As such, which of these two backgrounds is dominant depends on the tagging strategy, which itself would depend on the target jets. For the sake of this article, we will concentrate on the $t\bar{t}$ background.

The signals and the backgrounds are generated using \texttt{MadGraph} \cite{Alwall:2014hca} and an implementation of the benchmark in \texttt{Feynrules} \cite{Alloul:2013bka}. In particular, we simulated $10^6$ events for the background and $5\times 10^4$ events for the different signals. When neural networks are introduced later in the paper, half of the events are used for training and half for testing. Hadronization is handled via \texttt{PYTHIA}~8 \cite{Sjostrand:2007gs} using the Hidden Valley module \cite{Carloni:2011kk, Carloni:2010tw}. The parameters used for the Hidden Valley modules are included in Table~\ref{table:PHVSettings}. \texttt{Delphes} 3 \cite{deFavereau:2013fsa} is used for detector simulations using the ATLAS settings. We find that the results of this paper do not depend much on the details of the dark sector or the detector simulation.

\begin{table}[t]
  \begin{center}
	\begin{tabular}{llll}
	  \toprule
		Setting         & Value  & Setting        & Value  \\
		\cmrule
		NGauge          & 3      & Dark pion mass & 10 GeV \\
        nFlav           & 1      & Dark rho mass  & 21 GeV \\
        FSR             & On     & pTminFSR       & 11 GeV \\
        alphaOrder      & 1      & fragment       & On     \\
        Lambda          & 10 GeV & probVec        & 0.75   \\
        Dark quark mass & 10 GeV                           \\
      \bottomrule
	\end{tabular}
  \end{center}
	\caption{\texttt{PYTHIA}~8 Hidden Valley settings. All unstable hidden mesons are assumed to decay promptly. Hidden rho mesons are assumed to decay exclusively to dark pions. All other parameters are either specified in the text or left to their default values.} 
	\label{table:PHVSettings}
\end{table}

Many strategies could potentially be chosen to tag anomalous jets and the exact details are beyond the scope of this article. As such, we will simply assume a flat tagging efficiency for jets originating from a dark quark and fixed mistagging rates for other jets. Different rates will be considered. A few potential issues will be discussed in the conclusion.

All results from this paper are obtained by applying the following preliminary cuts. They are based on standard leptoquark search strategies. All jets and electrons considered must have a $p^T$ of at least 20~GeV and an $|\eta| < 2.5$. Every event is required to contain two jets that were tagged as semivisible jets and two opposite sign electrons. The electron pair is required to have an invariant mass $m_{ll} > 110$~GeV to suppress the $Z$ background. In a similar fashion to leptoquark searches, each electron can be paired with a jet and the pairing that minimizes the difference between the electron-jet pair invariant masses is selected. The mass asymmetry is then defined as
\begin{equation}\label{eq:masy}
 m_{\text{asym}} = \frac{m_{lj}^{\text{max}} - m_{lj}^{\text{min}}}{m_{lj}^{\text{max}} + m_{lj}^{\text{min}}},
\end{equation}
where $m_{lj}^{\text{max}}$ is the largest of the electron-jet pair invariant masses and $m_{lj}^{\text{min}}$ the smallest. We require $m_{\text{asym}} < 0.2$. All results will also include cuts on $E^T_{\text{miss}}$, which is the norm of $\mathbf{p}^T_{\text{miss}}$, and $S_T$, which is the scalar sum of the $p^T$ of the two electrons and two leading jets. These will differ from case to case and are presented where appropriate.

\section{Candidate event-level variables}\label{Sec:ELV}
In this section, we introduce a set of candidate event-level variables.

Before doing so, we introduce some relevant notation. Assume an event contains two electrons and two jets that were tagged as semivisible jets. Label the momenta of the jets tagged as semivisible as $\mathbf{p}_{d_1}$ and $\mathbf{p}_{d_2}$, ordered decreasingly by transverse magnitude. The momenta of the original dark quarks which led to these dark jets are referred to as $\mathbf{p}_{n_1}$ and $\mathbf{p}_{n_2}$ respectively. Similarly, label the momenta of the electrons as $\mathbf{p}_{e_1}$ and $\mathbf{p}_{e_2}$, ordered decreasingly by transverse magnitude, and the momenta of all jets as $\mathbf{p}_{j_i}$, also ordered decreasingly by transverse magnitude. The transverse components of vectors are denoted with a $T$ superscript, e.g. $\mathbf{p}_{d_1}^T$ is the transverse component of $\mathbf{p}_{d_1}$. The azimuthal angle of a vector $\mathbf{v}$ will be referred to as $\phi_{\mathbf{v}}$ and its pseudo-rapidity as $\eta_{\mathbf{v}}$. Finally, define $r_{\text{inv}}$ as the average fraction of the energy of the dark quark which initiated the semivisible jet that escapes the detector. In this section, we take the tagging efficiency of dark jets to be 1 and the mistagging rate of other jets to be negligible.

\subsection{Decomposition coefficients}\label{sSec:PC}
Given a sufficiently large $r_{\text{inv}}$, most of the $\mathbf{p}^T_{\text{miss}}$ in signals will generally come from the semivisible jets. This special fact means that, if the tagging worked properly, the main sources of $\mathbf{p}^T_{\text{miss}}$ is known. Each dark quark will generate a jet of transverse momentum $\sim(1 - r_{\text{inv}}) \mathbf{p}^T_{n_i}$ and a corresponding contribution to $\mathbf{p}_{\text{miss}}^T$ of $\sim r_{\text{inv}} \mathbf{p}_{n_i}^T$. One then expects a $\mathbf{p}_{\text{miss}}^T$ of
\begin{equation}\label{eq:vMETexp}
  \mathbf{p}_{\text{miss}}^T \sim \frac{r_{\text{inv}}}{1 - r_{\text{inv}}}\mathbf{p}_{d_1}^T + \frac{r_{\text{inv}}}{1 - r_{\text{inv}}}\mathbf{p}_{d_2}^T.
\end{equation}
Of course, this relation is not expected to hold exactly because of detector effects, other sources of $\mathbf{p}^T_{\text{miss}}$ and the fact that semivisible jets will contain varying amounts of stable particles. We will however demonstrate that it holds relatively well.

\begin{figure}[t!]
\begin{center}
	\includegraphics[width=0.75\textwidth]{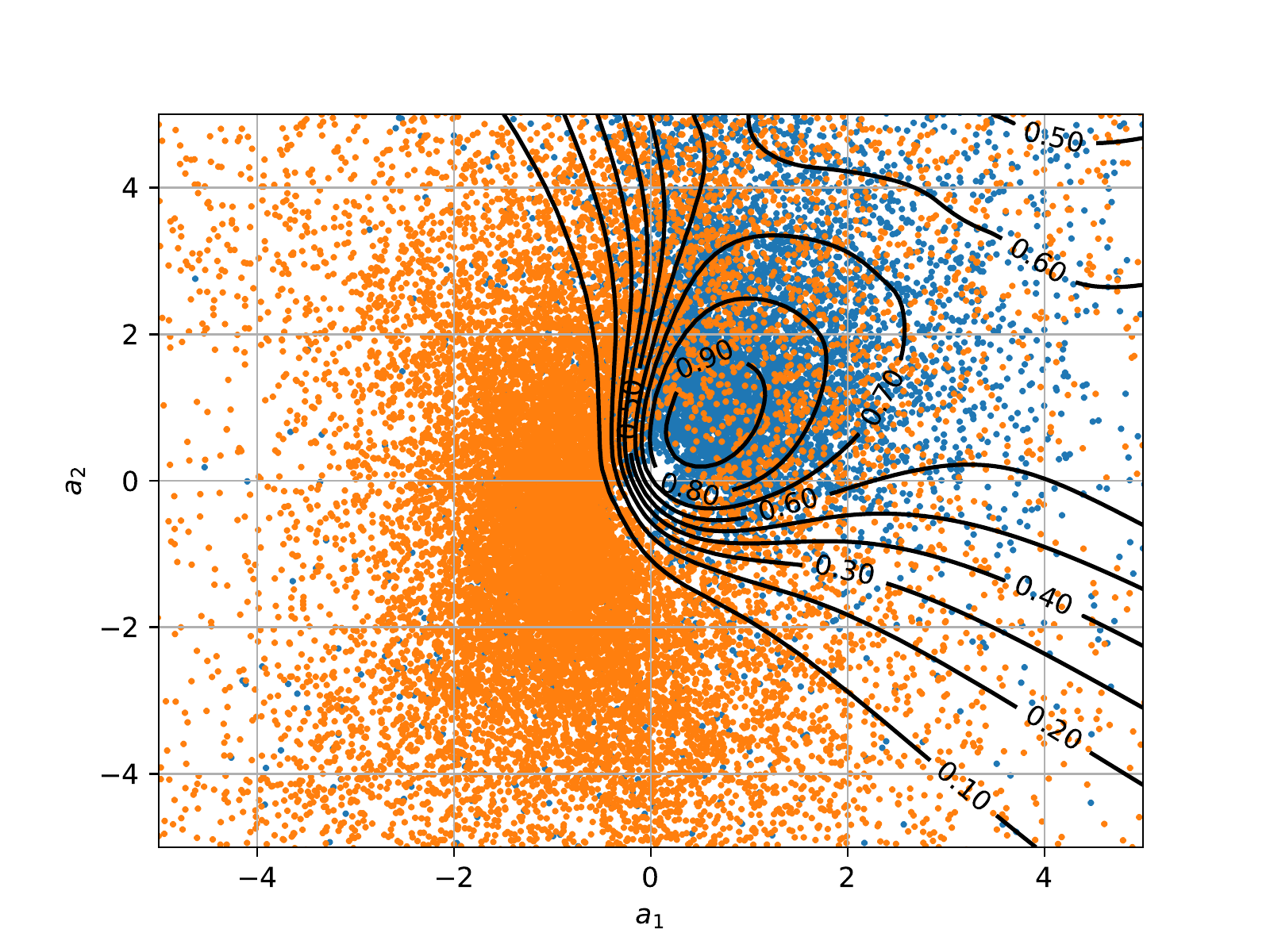}
\caption{Scatter plot of the background (orange) and signal (blue) with contours of the output of one of the neural networks for $r_{\mathrm{inv}} = 0.5$, a mediator mass $m_L=500$ GeV and minimal cuts  $\{E^T_{\text{miss}}, S_T\} > 100$ GeV.
\label{fig:NNstructurea1a2}}
\end{center}
\end{figure}

Considering this discussion, $\mathbf{p}_{\text{miss}}^T$ can be decomposed in terms of $\mathbf{p}^T_{d_1}$ and $\mathbf{p}^T_{d_2}$ as
\begin{equation}\label{eq:Mdecomposition}
  \mathbf{p}_{\text{miss}}^T = a_1 \mathbf{p}_{d_1}^T + a_2 \mathbf{p}_{d_2}^T.
\end{equation}
This gives
\begin{equation}\label{eq:a1a2}
  \begin{aligned}
    a_1 &= \frac{|\mathbf{p}_{d_2}^T|^2(\mathbf{p}_{d_1}^T\cdot\mathbf{p}_{\text{miss}}^T) - (\mathbf{p}_{d_1}^T\cdot\mathbf{p}_{d_2}^T)(\mathbf{p}_{d_2}^T\cdot\mathbf{p}_{\text{miss}}^T)}{|\mathbf{p}_{d_1}^T|^2|\mathbf{p}_{d_2}^T|^2 - (\mathbf{p}_{d_1}^T\cdot\mathbf{p}_{d_2}^T)^2},\\
    a_2 &= \frac{|\mathbf{p}_{d_1}^T|^2(\mathbf{p}_{d_2}^T\cdot\mathbf{p}_{\text{miss}}^T) - (\mathbf{p}_{d_1}^T\cdot\mathbf{p}_{d_2}^T)(\mathbf{p}_{d_1}^T\cdot\mathbf{p}_{\text{miss}}^T)}{|\mathbf{p}_{d_1}^T|^2|\mathbf{p}_{d_2}^T|^2 - (\mathbf{p}_{d_1}^T\cdot\mathbf{p}_{d_2}^T)^2}.
  \end{aligned}
\end{equation}
Comparing Eqs.~\eqref{eq:vMETexp} and \eqref{eq:Mdecomposition}, one expects the signal to give
\begin{equation}\label{eq:vDecomposition}
  a_1 \sim a_2 \sim \frac{r_{\text{inv}}}{1 - r_{\text{inv}}}.
\end{equation}
\begin{figure}[t!]
\begin{center}
 \begin{subfigure}{0.47\textwidth}
    \centering
    \caption{$a_1$}
    \includegraphics[width=1\textwidth]{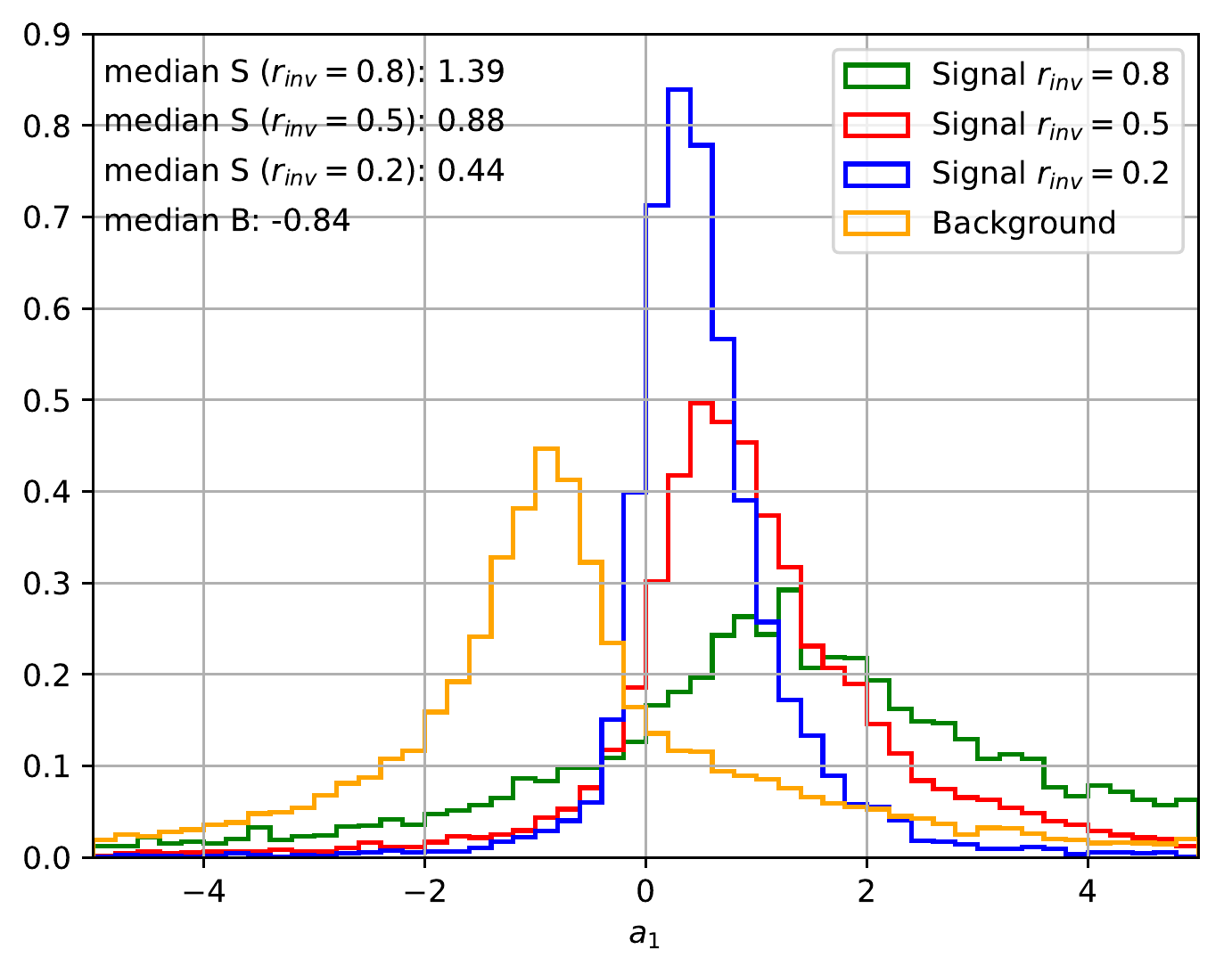}
    \label{fig:a1_0.2_500_100_100}
  \end{subfigure}
 \begin{subfigure}{0.47\textwidth}
    \centering
    \caption{$a_2$}
    \includegraphics[width=1\textwidth]{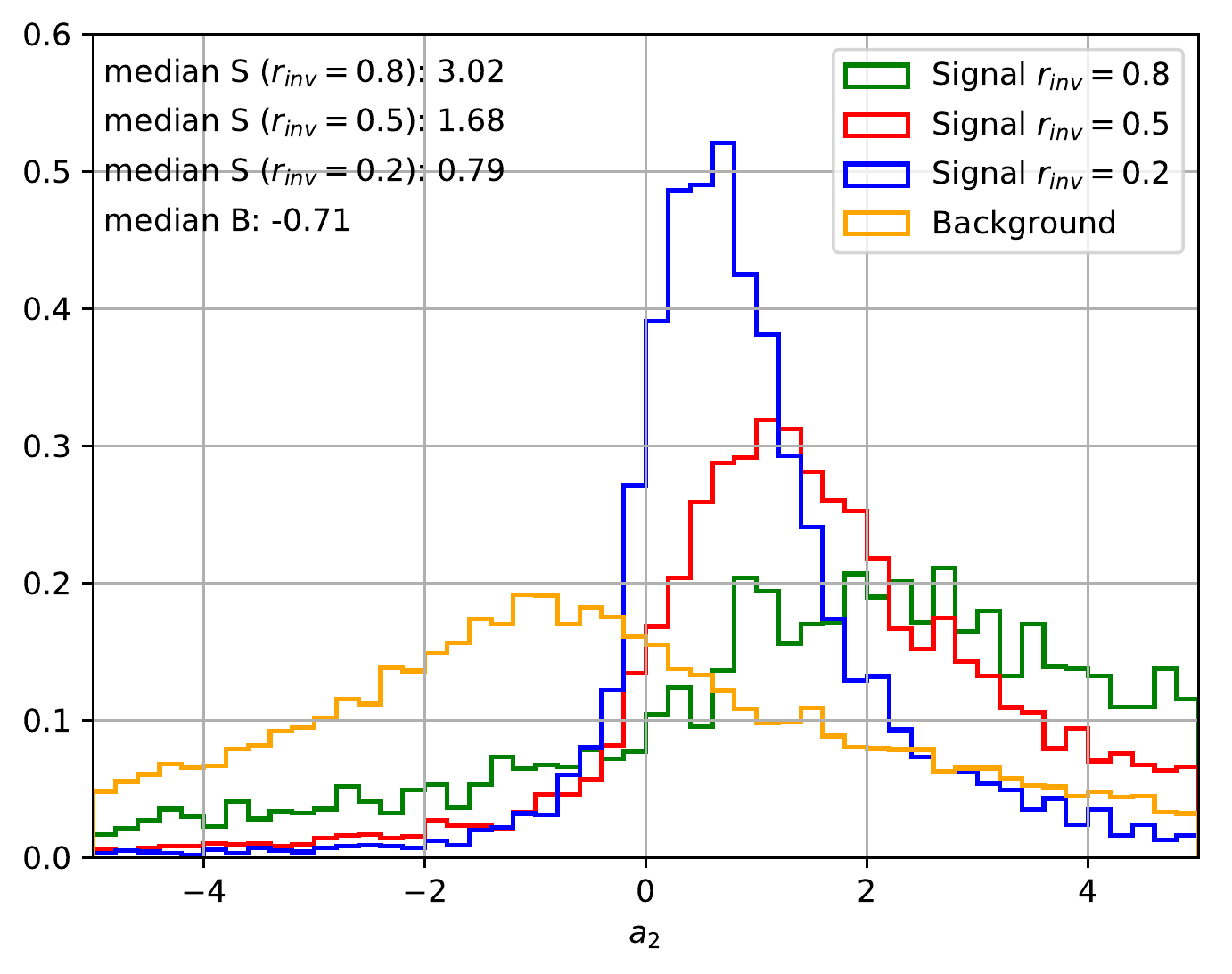}
    \label{fig:a2_0.2_500_100_100}
  \end{subfigure}
\caption{Distribution of (a) $a_1$ and (b) $a_2$ for background (orange) and signal (blue, red, green) events. Signal events were simulated for $r_{\mathrm{inv}}=0.2,\,0.5,\,0.8$, a mediator mass $m_L=500$ GeV and minimal cuts $\{E^T_{\text{miss}}, S_T\} > 100$ GeV.
\label{fig:a1_a2_dist}}
\end{center}
\end{figure}
The quantities $a_1$ and $a_2$ will serve as our first test statistics. An example of their distribution is shown in Fig.~\ref{fig:a1_a2_dist} for both a given signal and $t\bar{t}$. The signal events were simulated for a mediator mass of 500 GeV, different values of $r_{\mathrm{inv}}= 0.2$ (blue), $0.5$ (red)  and $0.8$ (green) and minimal cuts of $\{E^T_{\text{miss}}, S_T\} > 100$ GeV. As can be seen, the signal peaks around the expected value. There is a higher discrepancy at large $r_{\text{inv}}$, as the jets contain less visible hadrons which leads to a larger impact of statistical fluctuations and makes reconstructing the jet more difficult. The variables $a_1$ and $a_2$ will also prove to not be very correlated (see Fig.~\ref{fig:NNstructurea1a2}). In addition, the background peaks at negative values of both $a_1$ and $a_2$. The reason is the minimum requirement on $E^T_{\text{miss}}$. Indeed, the $E^T_{\text{miss}}$ for the background comes mainly from the neutrinos. If the $E^T_{\text{miss}}$ cut is going to be satisfied, the neutrinos must have similar azimuthal angles. Conservation of transverse momentum means that the other objects are more likely to be in the opposite direction in the transverse plane. This in turn means that the jets are more likely to be in the opposite direction of $\mathbf{p}^T_{\text{miss}}$, hence why the $a_i$ tend to be negative.

For comparison purposes, it will be necessary to combine $a_1$ and $a_2$ into a single test statistic. The optimal cut on these variables is however not trivial and we train a neural network to generate the optimal selection. The architecture of the neural network is shown in Table~\ref{sFig:NN1}. It is trained with fully labelled data to return 1 for a signal and 0 for the $t\bar{t}$ background. The different hyperparameters are shown in Table~\ref{sFig:HP1}. This neural network is implemented via the \textsc{Python} deep learning library \textsc{Keras}~\cite{chollet2015keras} with the \textsc{TensorFlow} backend~\cite{Abadi:2016kic} using the \textsc{Adam} optimizer~\cite{Kingma:2014vow}. To minimize the variance of the predictions associated with the weight and bias initialization, model averaging is performed over 30 neural networks. Contours of the output of an example neural network are shown in Fig.~\ref{fig:NNstructurea1a2} for $r_{\mathrm{inv}}=0.5$ and $m_L=500$ GeV. As can be seen, the neural network learns well how to identify the region where most signals belong. Considering the simple nature of the results, the variables $a_1$ and $a_2$ could be combined in a boosted decision tree or a diagonal cut in the $a_1$-$a_2$ plane could be made. Both approaches would not affect much the results as long as they are properly implemented.

\begin{table}[t]
	\begin{subfigure}{.5\linewidth}
		{\footnotesize
			\setlength\tabcolsep{5pt}
			\begin{center}
				\begin{tabular}{ll}
					\toprule
					Layer                  & Parameters \\
					\cmrule
					Input layer            & 2 bins                              \\
					Dense layer 1          & $\#$ nodes = 100                    \\
					& Activation: Elu                                            \\
					Dense layer 2          & $\#$ nodes = 50                     \\
					& Activation: Elu                                            \\
					Dense layer 3          & $\#$ nodes = 25                     \\
					& Activation: Elu                                            \\
					Dense layer 4          & $\#$ of nodes = 1                   \\
					& Activation: Sigmoid                                        \\
					\bottomrule
				\end{tabular}
			\end{center}
		}
		\caption{}
        \label{sFig:NN1}
	\end{subfigure}
	\begin{subfigure}{.5\linewidth}
		{\footnotesize
			\setlength\tabcolsep{5pt}
			\begin{center}
				\begin{tabular}{ll}
					\toprule
					Setting                   & Choice \\
					\cmrule
					Optimizer                 & \textsc{Adam}        \\
					Loss function             & Binary cross entropy \\
					Validation split          & 0.2                  \\
					Batch size                & 1000                 \\
					$\#$ epochs               & 200                  \\
                    Learning rate             & $10^{-3}$            \\
                    Callback                  & Smallest validation  \\
					& loss function        \\
					\bottomrule
				\end{tabular}
			\end{center}
		}
		\caption{}
        \label{sFig:HP1}
	\end{subfigure}
	\caption{(a) Structure of the neural network for the projection coefficients $a_1$ and $a_2$. (b) Training parameters. All parameters not specified in these tables are left at their default \textsc{Keras} values.} 
	\label{table:NNstructurea1a2}
\end{table}

\subsection{Azimuthal angle difference}\label{sSec:AAD}
\begin{figure}[t!]
\begin{center}
 \begin{subfigure}{0.47\textwidth}
    \centering
    \caption{$\Delta \phi$}
    \includegraphics[width=1\textwidth]{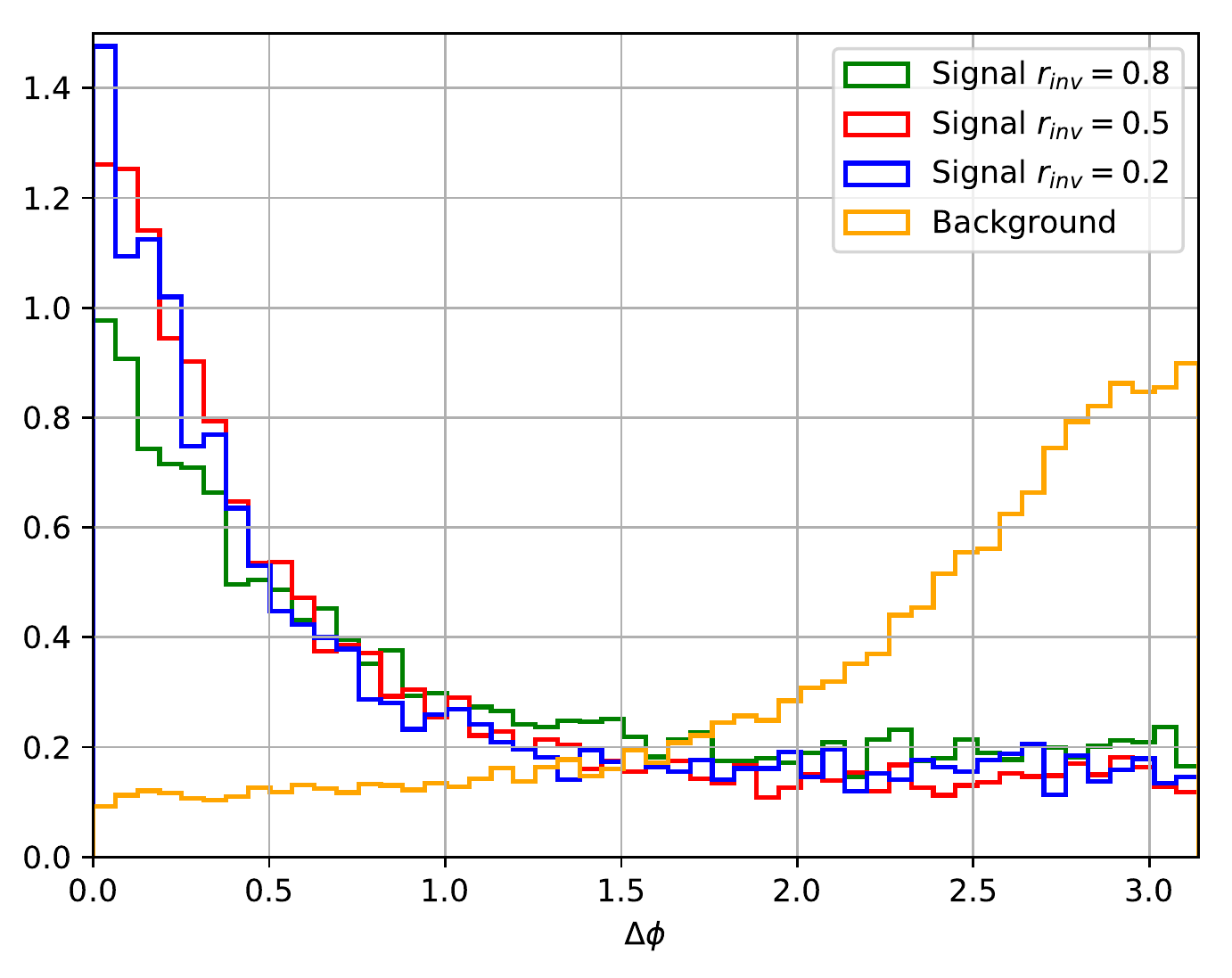}
    \label{sfig:Deltaphi}
  \end{subfigure}
  ~
  \begin{subfigure}{0.47\textwidth}
    \centering
    \caption{$\Delta \phi_{\text{CLLM}}$}
    \includegraphics[width=1\textwidth]{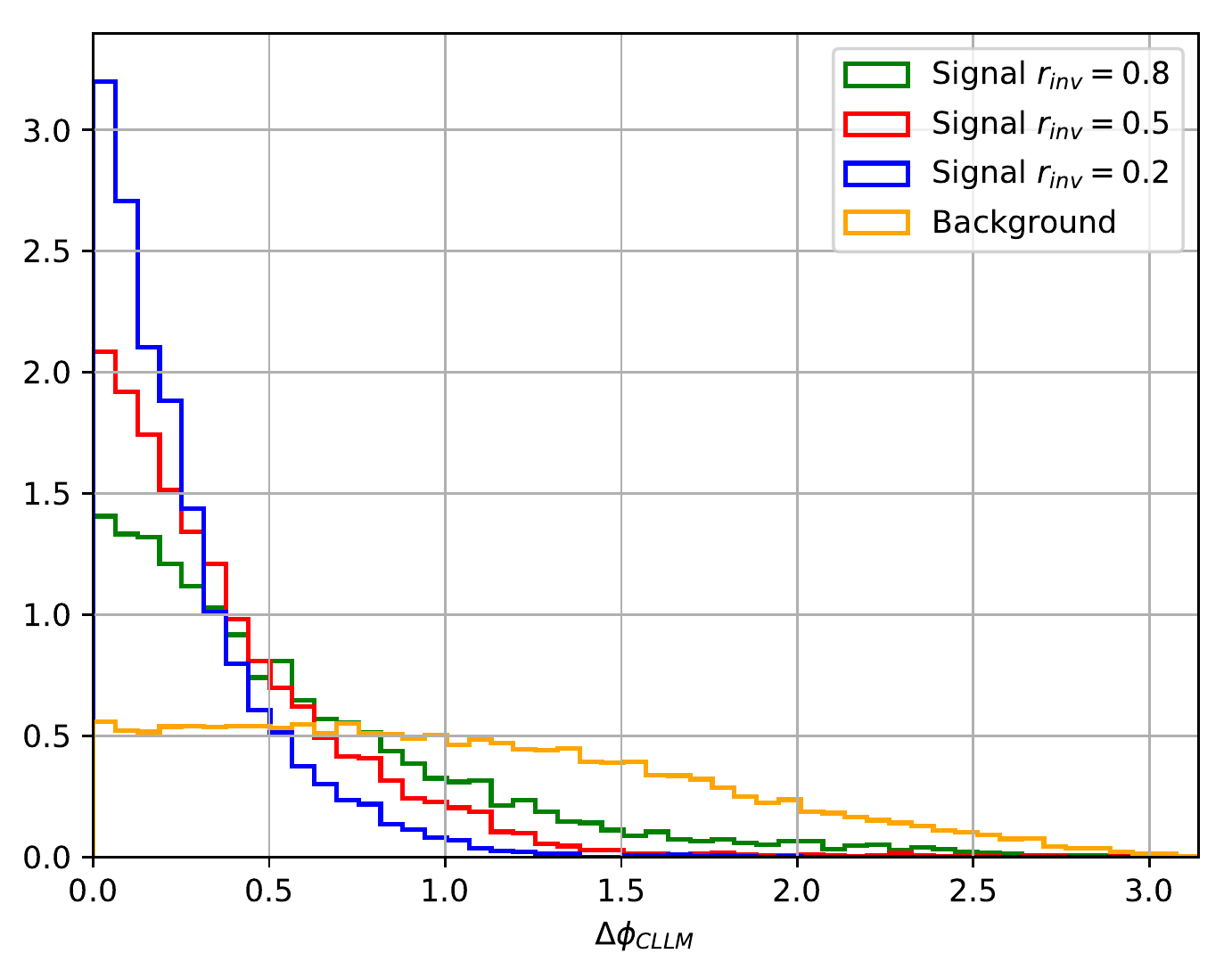}
    \label{sfig:DeltaphiCLLM}
  \end{subfigure}
\caption{Distributions of the azimuthal angle difference variable (a) $\Delta \phi$ and (b) the minimal angular difference $\Delta \phi_{\text{CLLM}}$, for background (orange) and signal (blue, red, green) events. Signal events were simulated for $r_{\mathrm{inv}}=0.2,\,0.5$ and $0.8$, a mediator mass $m_L=500$ GeV and minimal cuts  $\{E^T_{\text{miss}}, S_T\} > 100$~GeV.
\label{fig:deltaphi_deltaphiCLLM}}
\end{center}
\end{figure}

A simple way to encode the logic of the last section into a single variable is to exploit the fact that the azimuthal direction of $\mathbf{p}^T_{\text{miss}}$ should be the same as that of the sum of the transverse momenta of the jets tagged as anomalous, i.e.
\begin{equation}\label{eq:pD}
  \mathbf{p}_{D}^T = \mathbf{p}_{d_1}^T  + \mathbf{p}_{d_2}^T.
\end{equation}
Therefore, the quantity
\begin{equation}\label{eq:DeltaPhi}
  \Delta \phi = \left|\phi_{\mathbf{p}_{D}^T} - \phi_{\mathbf{p}_{\text{miss}}^T}\right|
\end{equation}
should peak at 0 for the signal. As can be seen in Fig.~\ref{sfig:Deltaphi}, where we show the distribution of this quantity for a mediator mass of 500 GeV and different values of $r_{\mathrm{inv}}= 0.2$ (blue), $0.5$ (red) and $0.8$ (green), this is indeed the case. The background (orange), on the other hand, peaks at $\pi$ for the same reason that the background peaks at negative values of $a_1$ and $a_2$. For both signal and background we used minimal cuts of $\{E^T_{\text{miss}}, S_T\} > 100$ GeV. This variable is similar to one defined in Ref.~\cite{Beauchesne:2017yhh}, albeit that paper tried to identify the semivisible jets by looking at the invariant masses of jet and electron pairs. The main drawback of $\Delta \phi$ is that some information is lost in the process, as the prediction on the magnitude of $\mathbf{p}^T_{\text{miss}}$ is lost.

\subsection{Minimal angular difference}\label{sSec:MAD}
Refs.~\cite{Cohen:2015toa, Cohen:2017pzm} considered the possibility of simply comparing the azimuthal direction of $\mathbf{p}^T_{\text{miss}}$ with that of the closest sufficiently energetic jet. As such, we can define as in Ref.~\cite{Cohen:2017pzm}
\begin{equation}\label{eq:CLLM}
  \Delta \phi_{\text{CLLM}} = \min_{i \leq 4}\left\{\left|\phi_{\mathbf{p}_{j_i}^T}  - \phi_{\mathbf{p}_{\text{miss}}^T}\right|\right\},
\end{equation}
where the inequality means that only the four leading jets are considered. The study of Ref.~\cite{Cohen:2017pzm} was focused on dijet production and as such this variable is not expected to be optimal for the current benchmark. Distributions of this variable are shown in Fig.~\ref{sfig:DeltaphiCLLM} for both a signal with $r_{\mathrm{inv}}=0.2$ (blue), $\,0.5$ (red) and $0.8$ (green), a mediator mass $m_L=500$ GeV and the $\bar{t}t$ background (orange). We used the minimal cuts $\{E^T_{\text{miss}}, S_T\} > 100$~GeV for both signal and background.

\subsection{Comparison with fully supervised neural network}\label{sSec:NN}
\begin{figure}[t!]
\begin{center}
	\includegraphics[width=0.5\textwidth]{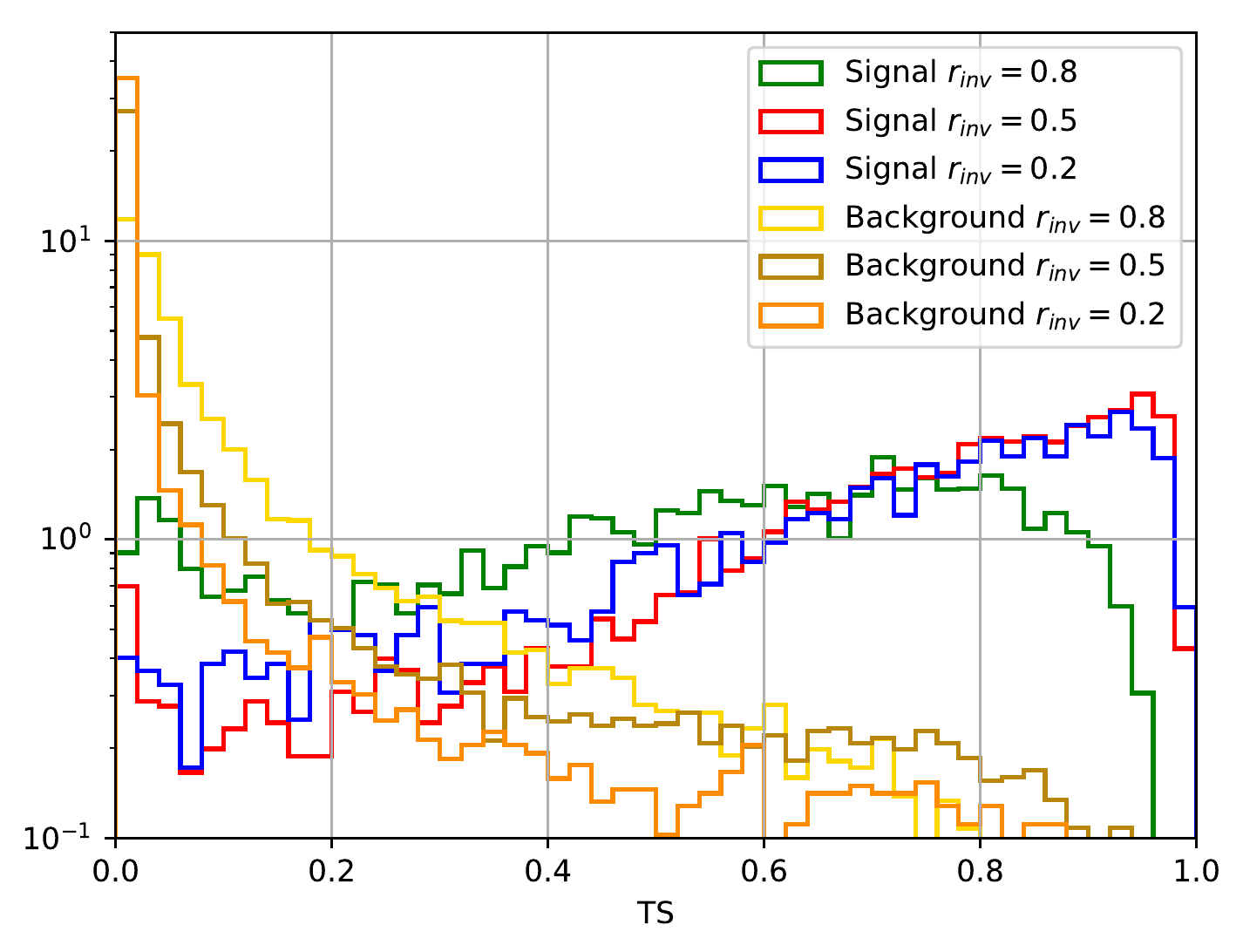}
\caption{Output distributions of the fully supervised neural network with inputs $x$ (see Eq. \eqref{eq:NNinputs}) for background (orange, brown, yellow) and signal (blue, red, green) events. Signal events were simulated for $r_{\mathrm{inv}}=0.2,\,0.5$ and $0.8$, a mediator mass $m_L=500$ GeV and minimal cuts  $\{E^T_{\text{miss}}, S_T\} > 100$ GeV.
\label{fig:FSNNall}}
\end{center}
\end{figure}

Finally, we consider a fully supervised neural network trained on the angular information of the different objects. This is not meant to be used as part of an actual analysis, as training a neural network on simulation always introduces the risk of the neural network becoming sensitive to simulation artefacts. Instead, it is meant to check how close the other methods are to the ideal scenario, as a classifier neural network trained with full supervision usually gives close to optimal results. The inputs of the neural network are
\begin{equation}\label{eq:NNinputs}
  x = \{\phi_{\mathbf{p}_{d_1}}, \phi_{\mathbf{p}_{d_2}}, \eta_{\mathbf{p}_{d_1}}, \eta_{\mathbf{p}_{d_1}}, \phi_{\mathbf{p}_{\text{miss}}^T}\}.
\end{equation}
The architecture is almost identical to that of Table~\ref{table:NNstructurea1a2}, the only difference being that the number of inputs is now 5. An example of the test statistic distributions is shown in Fig.~\ref{fig:FSNNall} for both a given signal and the $t\bar{t}$ background. The signals are simulated for a mediator mass $m_L=500$ GeV and different values of $r_{\mathrm{inv}} = 0.2$ (blue), $0.5$ (red) and $0.8$. Both signals and backgrounds make use of minimal cuts on $\{E^T_{\text{miss}}, S_T\} > 100$ GeV. Notice that in Fig.~\ref{fig:FSNNall} we show the three different background histograms corresponding to the signal with different $r_{\mathrm{inv}}$. This is motivated by the fact that different neural networks were trained for each value of $r_{\mathrm{inv}}$, leading to the definition of three different test statistic.

\section{Results}\label{Sec:Results}
Having defined the candidate event-level variables, they can now be compared with each other and the increased significance studied.

\subsection{Comparison}\label{sSec:Comparison}

\begin{figure}[t!]
\begin{center}
 \begin{subfigure}{0.47\textwidth}
    \centering
    \caption{$m_L = 500$ GeV, $r_{\mathrm{inv}}=0.2$}
    \includegraphics[width=1\textwidth]{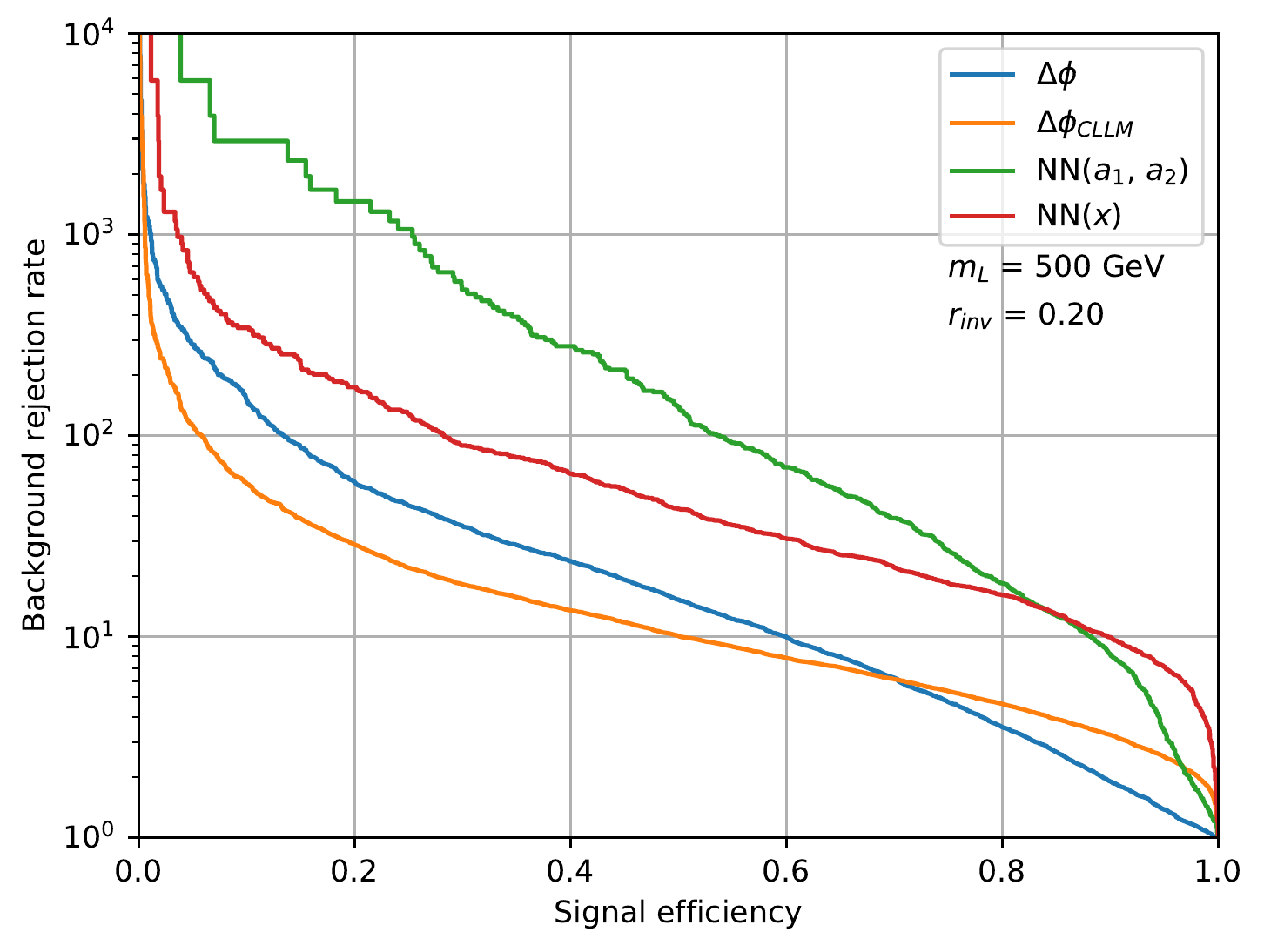}
    \label{fig:ROC_0.2_500_100_100}
  \end{subfigure}
  ~
  \begin{subfigure}{0.47\textwidth}
    \centering
    \caption{$m_L = 500$ GeV, $r_{\mathrm{inv}}=0.4$}
    \includegraphics[width=1\textwidth]{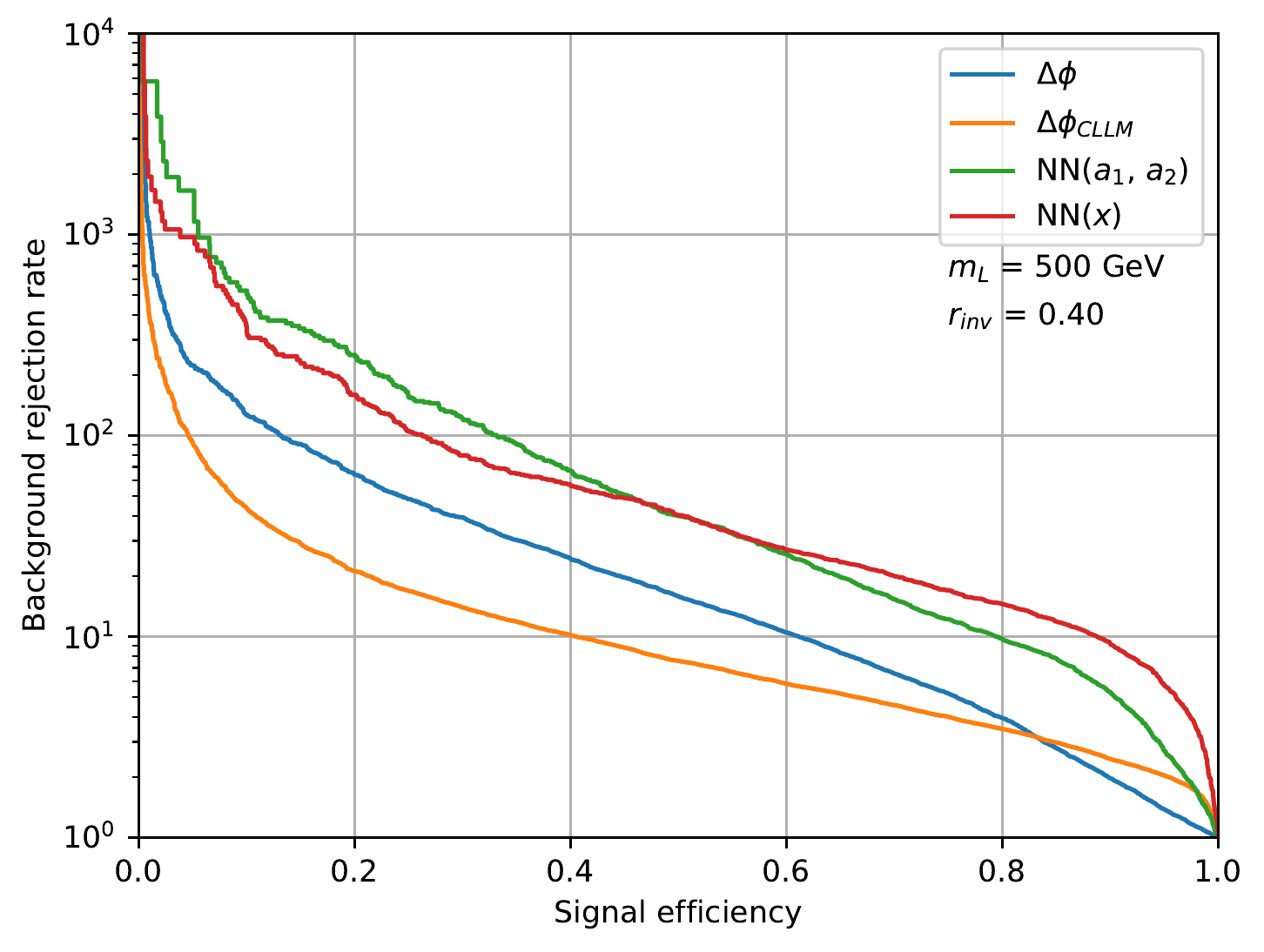}
    \label{fig:ROC_0.4_500_100_100}
  \end{subfigure}
  ~
  \begin{subfigure}{0.47\textwidth}
    \centering
    \caption{$m_L = 500$ GeV, $r_{\mathrm{inv}}=0.6$}
    \includegraphics[width=1\textwidth]{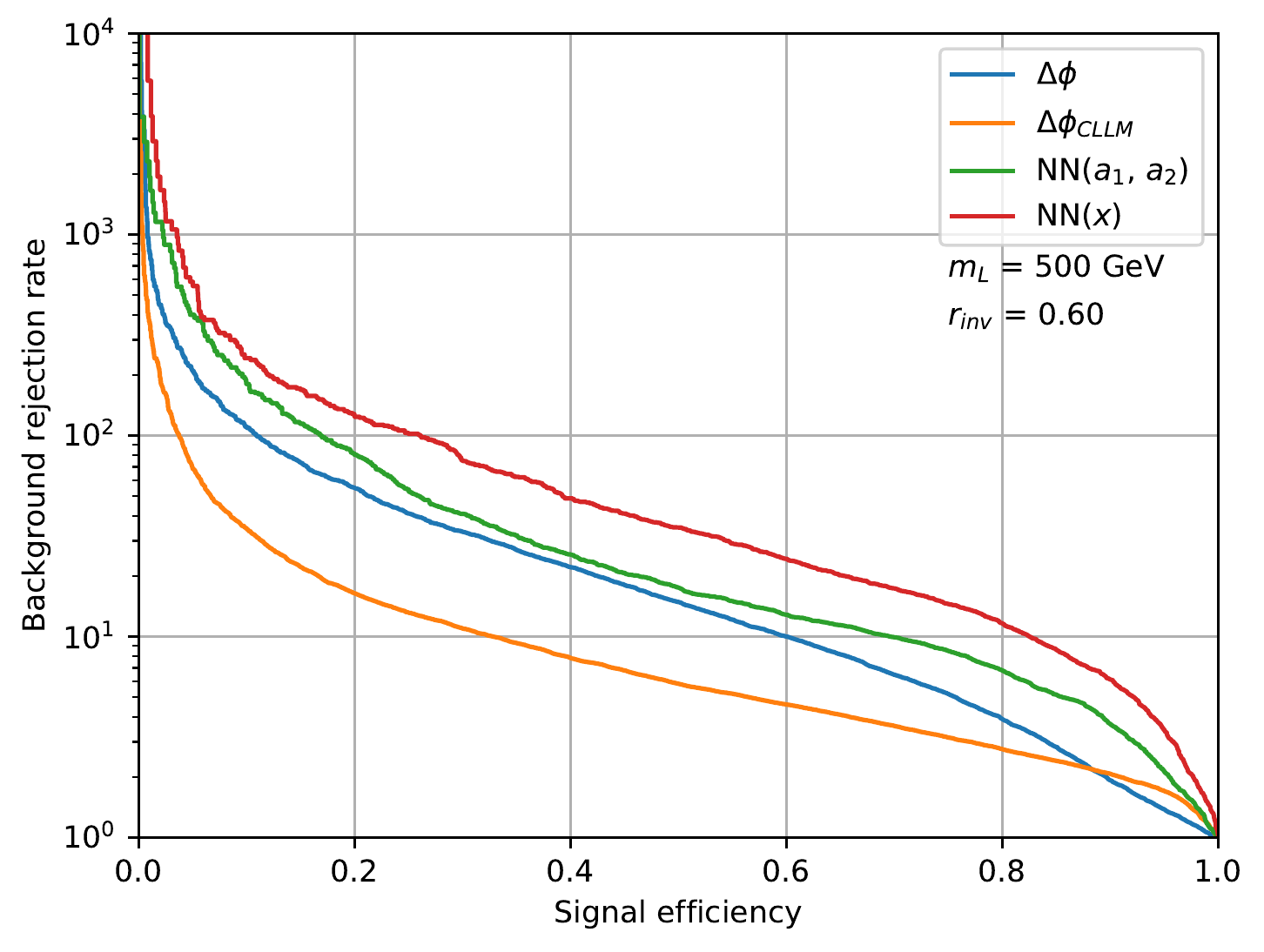}
    \label{fig:ROC_0.6_500_100_100}
  \end{subfigure}
  ~
  \begin{subfigure}{0.47\textwidth}
    \centering
    \caption{$m_L = 500$ GeV, $r_{\mathrm{inv}}=0.8$}
    \includegraphics[width=1\textwidth]{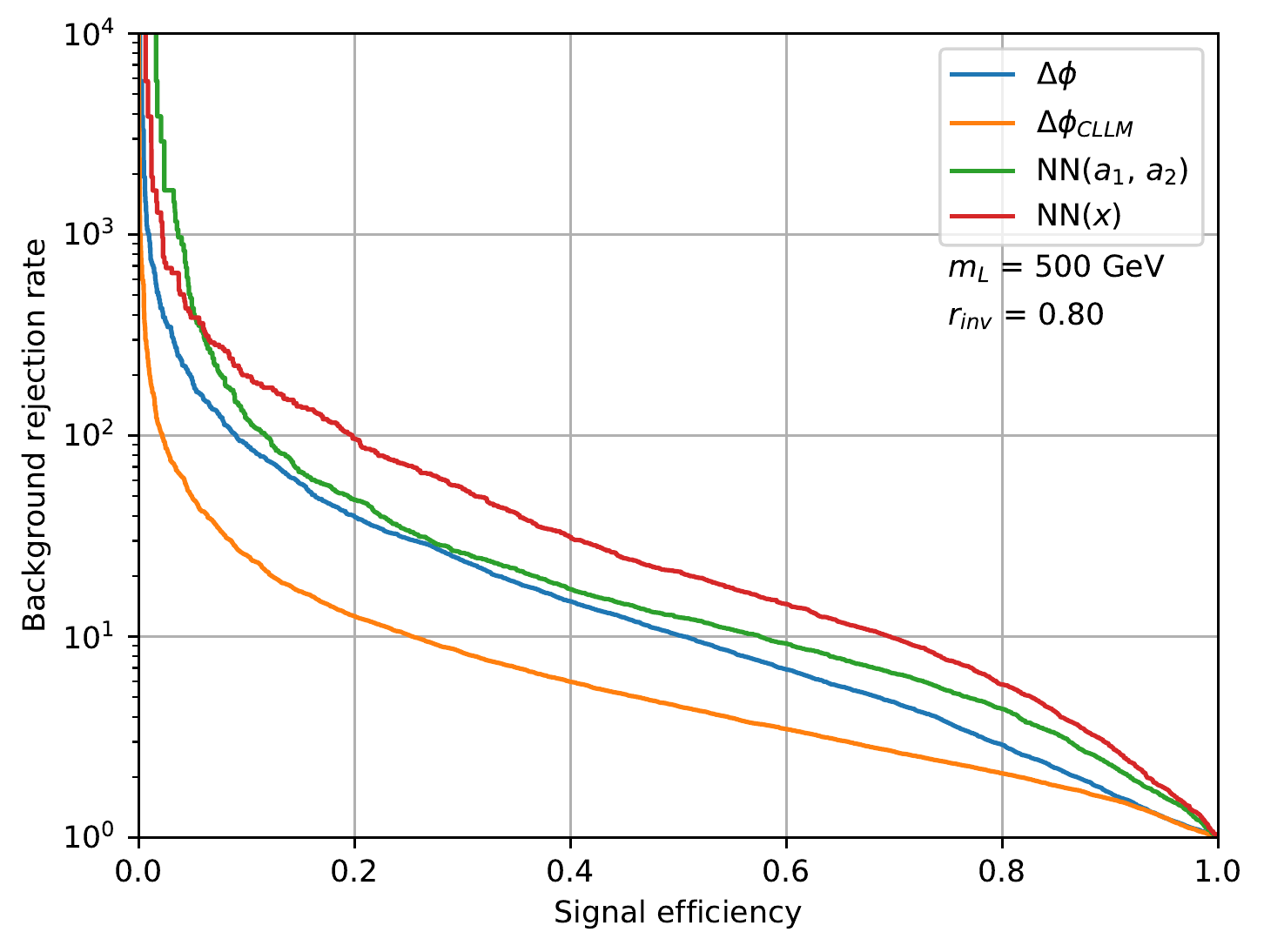}
    \label{fig:ROC_0.8_500_100_100}
  \end{subfigure}
  ~
\caption{ROC curves for $\Delta \phi$ (blue), $\Delta \phi_{CLLM}$ (orange) and for the output of the neural networks that have as input the variables $a_1$ and $a_2$ (green) or the set $x$ in Eq.~\eqref{eq:NNinputs} (red). The four plots differs for different values of $r_{\mathrm{inv}} = 0.2,\,0.4,\,0.6$ and $0.8$, a mediator mass $m_L=500$ GeV and minimal cuts  $\{E^T_{\text{miss}}, S_T\} > 100$ GeV. 
\label{fig:ROC}}
\end{center}
\end{figure}

 Receiver Operating Characteristic (ROC) curves are shown in Fig.~\ref{fig:ROC} for $\Delta \phi$, $\Delta \phi_{CLLM}$ and for the output of the neural networks that have as input the variables $a_1$ and $a_2$ or the set $x$ of Eq.~\eqref{eq:NNinputs}. The signals were simulated for $m_L=500$ GeV and $r_{\mathrm{inv}} = 0.2,\,0.4,\,0.6$ and $0.8$, using a tagging efficiency of $\sim$1, a very small background mistagging rate and minimal cuts on $\{E^T_{\mathrm{miss}}, S_T\} >100$ GeV. As can be seen, the best results are generally obtained for the combination of the projection coefficients. The fact that this beats the fully supervised neural network is because these variables are not fully independent of the momenta of the jets and the $E^T_{\text{miss}}$. This information is simply not provided to the fully supervised neural network. When the fully supervised neural network does beat the $a_1$ and $a_2$ variables, it barely does so. The azimuthal angle difference $\Delta \phi$ performs only slightly worse than the fully supervised neural network, but still provides a relatively large discriminating power. Finally, the variable $\Delta \phi_{\text{CLLM}}$ of Ref.~\cite{Cohen:2017pzm} performs the worst. This was expected as this variable was designed for dijet events and does not utilize the tagging information.

\subsection{Increased significance}\label{sSec:IS}
We now study how much the new event-level variables can increase our ability to discover dark confining sectors using the increased significance. This is defined as the signal efficiency $\epsilon_s$ of a given cut divided by the square root of its background efficiency $\epsilon_B$, i.e. $\epsilon_S/\sqrt{\epsilon_B}$. We scan over the mass of the mediator $m_X$ and $r_{\text{inv}}$. To take into account that multiple signal regions would normally be defined, we use different minimal values of the $E^T_{\text{miss}}$ and $S_T$. All results are presented for the combination of the decomposition coefficients $a_1$ and $a_2$.

\begin{figure}[t!]
\begin{center}
 \begin{subfigure}{0.47\textwidth}
    \centering
    \caption{$S_T > 100$ GeV, $E^T_\text{miss} > 100$ GeV}
    \includegraphics[width=1\textwidth]{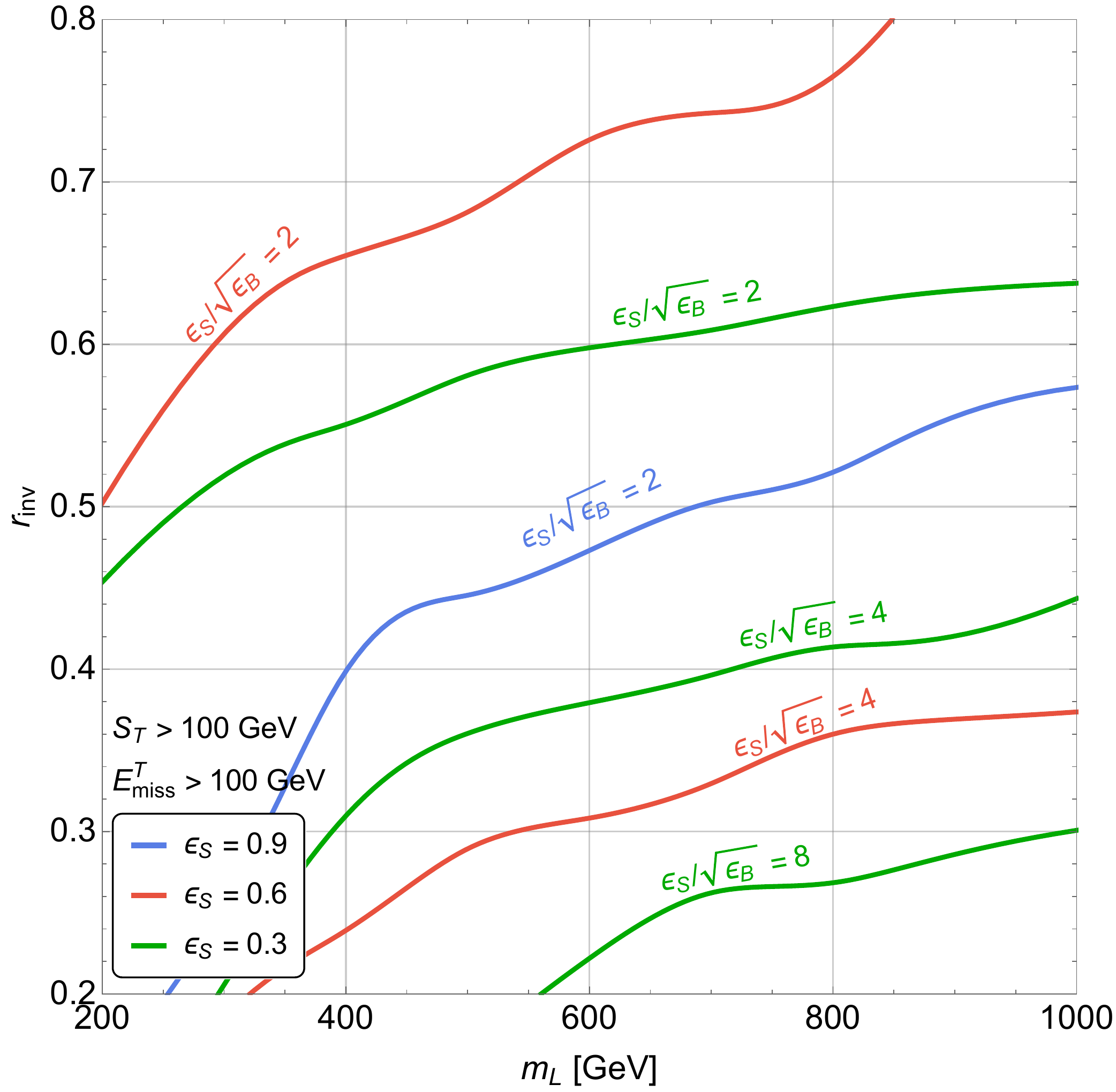}
  \end{subfigure}
  ~
  \begin{subfigure}{0.47\textwidth}
    \centering
    \caption{$S_T > 100$ GeV, $E^T_\text{miss} > 150$ GeV}
    \includegraphics[width=1\textwidth]{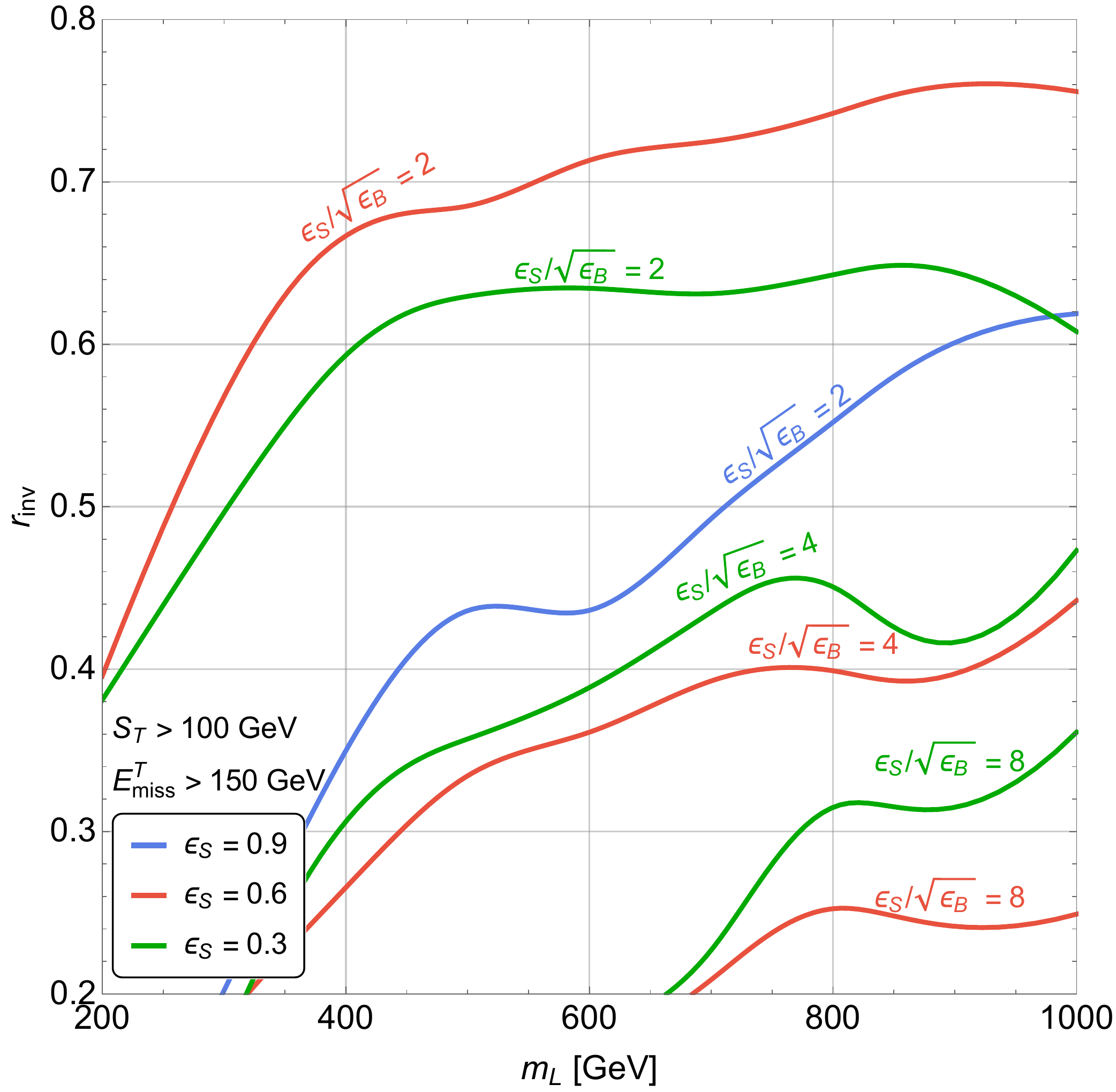}
  \end{subfigure}
\caption{Increased significance for $\epsilon_s = 0.3,\,0.6,\,0.9$ in the $m_L$ vs $r_{\mathrm{inv}}$ plane for different cuts in $E^T_{\text{miss}}$ and $S_T$ and a dark jet tagging efficiency of 1 and negligible QCD jet (or b-jet) mistagging rates.
\label{fig:incr_sig_DJt1}}
\end{center}
\end{figure}
We show in Fig. \ref{fig:incr_sig_DJt1} the increased significance for a high tagging efficiency and low background mistagging rate. This is done for three different values of $\epsilon_s = 0.3,\,0.6,\,0.9$. As can be seen, the increased significance can reach values close to 8. The variables $a_1$ and $a_2$ are however not expected to work appropriately when $r_{\text{inv}}$ is close to either 0 or 1. For $r_{\text{inv}}$ close to 0, there are simply no stable dark hadrons that escape the detector, the $\mathbf{p}^T_{\text{miss}}$ originates from other sources and the correlation is gone. For $r_{\text{inv}}$ close to 1, the dark jets are almost completely invisible and thus might not even pass the selection cuts. We therefore omit from our scans the regions $r_{\text{inv}} < 0.2$ and $r_{\text{inv}} > 0.8$.

\begin{figure}[t!]
\begin{center}
 \begin{subfigure}{0.47\textwidth}
    \centering
    \caption{$S_T > 100$ GeV, $E^T_\text{miss} > 100$ GeV}
    \includegraphics[width=1\textwidth]{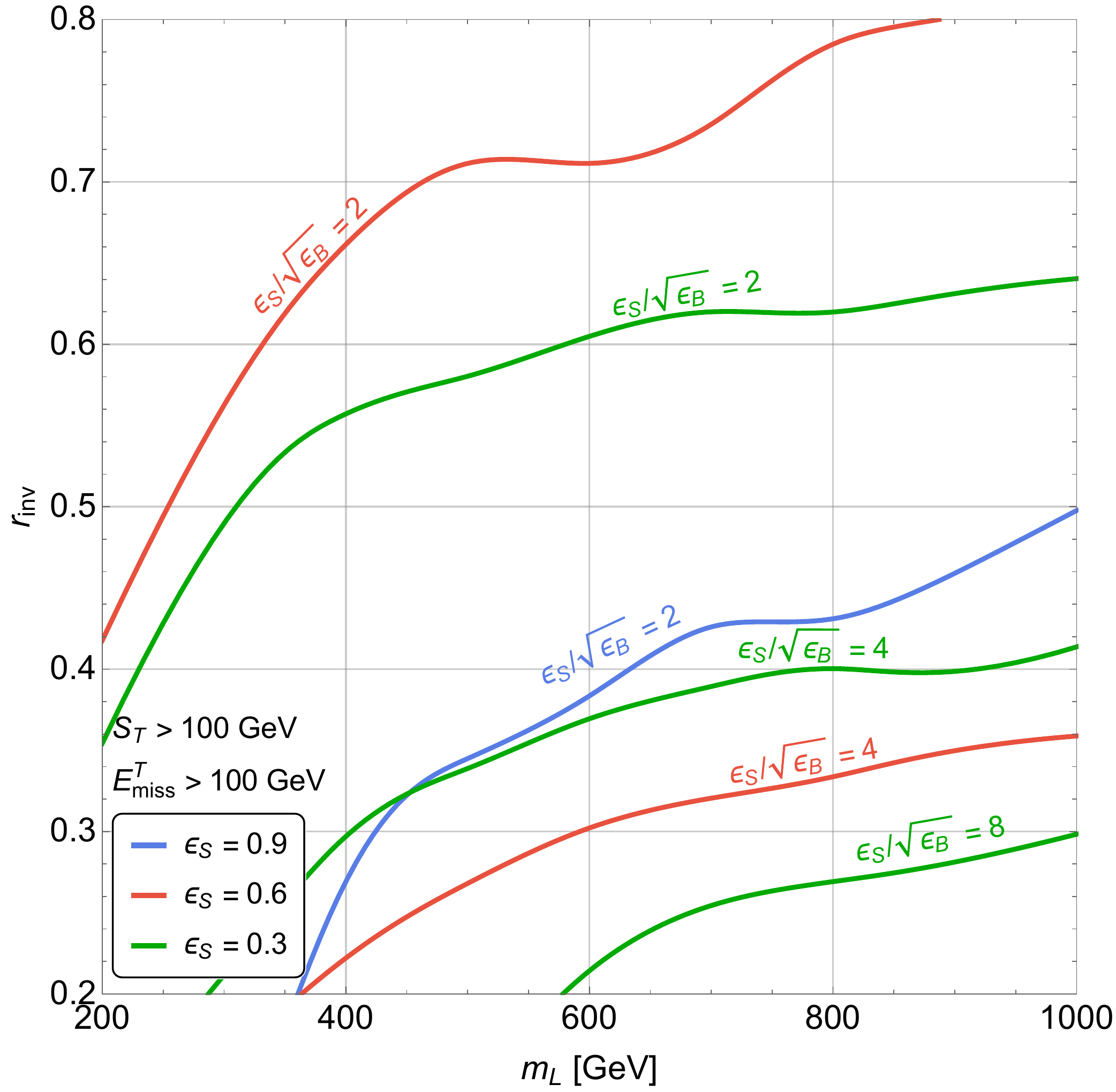}
  \end{subfigure}
  ~
  \begin{subfigure}{0.47\textwidth}
    \centering
    \caption{$S_T > 100$ GeV, $E^T_\text{miss} > 150$ GeV}
    \includegraphics[width=1\textwidth]{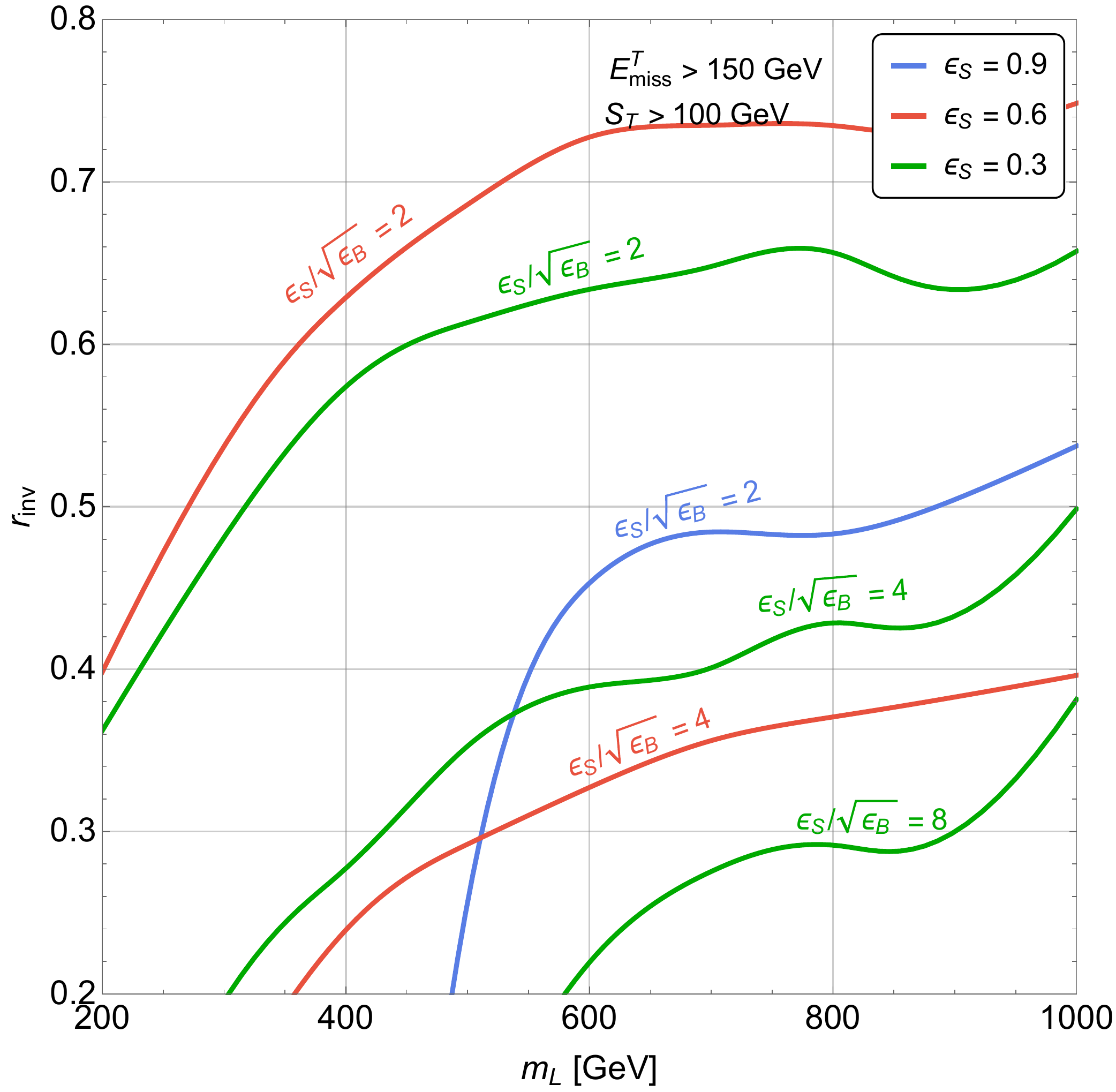}
  \end{subfigure}
\caption{Increased significance for $\epsilon_s = 0.3,\,0.6,\,0.9$ in the $m_L$ vs $r_{\mathrm{inv}}$ plane for different cuts in $E^T_{\text{miss}}$ and $S_T$ and a dark jet tagging efficiency of 0.8 and a QCD jet (or b-jet) mistagging rate of 0.05.
\label{fig:incr_sig_DJt0.8}}
\end{center}
\end{figure}

We show in Fig.~\ref{fig:incr_sig_DJt0.8} the increased significance for a considerably worse tagging efficiencies of 0.8 and a QCD jet (or b-jet) mistagging rate of 0.05. As can be seen, the decomposition coefficients can still work very well despite imperfect tagging.

\section{Conclusion}\label{Sec:Conclusion}
Semivisible jets are a common signature of many confining dark sectors and consist of jets of visible hadrons intermixed with stable particles. Their special nature leads to unique relations between the direction of the jets and $\mathbf{p}^T_{\text{miss}}$. Over the past few years, there has been much progress in developing techniques to tag anomalous jets. The goal of this paper was therefore to use the ability to tag semivisible jets to define new event-level variables to improve the reach of searches for confining dark sectors.

The end result was that such variables can indeed be designed and lead to a sizable suppression of the background. The best results were obtained with the decomposition of $\mathbf{p}^T_{\text{miss}}$ in terms of the jets tagged as anomalous. The azimuthal direction of the sum of the momenta of the jets tagged as anomalous was also shown to give good results. The new variables were shown to give better results than previous event-level variables. We also demonstrated that these variables can easily increase the significance of an excess by almost an order of magnitude. Finally, the variables were shown to still work well despite imperfect tagging.

We conclude this paper by mentioning that, like any other variables, there are limitations to the event-level variables defined in this paper and that there might be situations in which these variables would not be optimal.

First, the variables would not work in the presence of too many additional stable particles in the signal. For example, the mediator $X$ could interact with left-handed leptons. The signal would then often include neutrinos which would contribute significantly to $\mathbf{p}^T_{\text{miss}}$. When the mediator decays to neutrinos, one would then be forced to resort to other kinematic variables, though our variables could still be applied when both mediators decay to charged leptons.

Second, certain techniques for tagging anomalous jets could reduce the effectiveness of the variables. Consider the following example. In many backgrounds, $\mathbf{p}^T_{\text{miss}}$ comes for mismeasurements of the energy of the components of a jet or some components outright being omitted. Assume this is especially the case for a given jet. This jet would therefore appear somewhat peculiar and some suboptimal tagger could therefore tag it as anomalous. Since it should be the main source of $\mathbf{p}^T_{\text{miss}}$, the jet in question should be nearly collinear with $\mathbf{p}^T_{\text{miss}}$ and our variables would be less efficient. The solution to this problem would be to design the tagger such that jets which seem to be missing too many components or have their energy misrecontructed not be tagged as anomalous. The exact details are however beyond the scope of this paper.      

Finally, we considered in this paper a benchmark that leads to jets with a considerably large multiplicity of dark hadrons. This meant that the variation on the fraction of a semivisible jets that goes to invisible is relatively small. For dark sectors with lower multiplicity, the distributions of the different variables would be wider and some of their power would be lost.

\acknowledgments
This research was supported in part by the Israel Science Foundation (grant no.\ 780/17) and the United States - Israel Binational Science Foundation (grant no.\  2018257). This work was supported by the Ministry of Science and Technology, National Center for Theoretical Sciences of Taiwan. GGdC is supported by the INFN Iniziativa Specifica Theoretical Astroparticle Physics (TAsP) and by the Frascati National Laboratories (LNF) through a Cabibbo Fellowship call 2019. We furthermore acknowledge the use of the High Performance Computing cluster at CNAF, and thank F. Minarini and A. Falabella for providing technical support.

\bibliography{biblio}
\bibliographystyle{utphys}

\end{document}